**Electrochemical functionalization of graphene rationalized by Hammett and Swain-Lupton parameters**

L. Chen[1,2], H. Li[1], C. Li[1,2], X. Zhang[1,2], Q. Dai[1,2], M. Alhashmi[1,2], J. Tong[1,2], E. Hoenig[1,2], Y. Li[3,4], A. Campbell[1,2], K. Hu[1,2], Y.-C. Soong[1,2], C. Sevik[5], L. F. Vega[3,4], F. M. Peeters[5,6,7], M. Lozada-Hidalgo[1,2]

[1] Department of Physics and Astronomy, The University of Manchester, Manchester M13 9PL, UK

[2] National Graphene Institute, The University of Manchester, Manchester M13 9PL, UK

[3]Research and Innovation Center on $CO_2$ and Hydrogen (RICH Center) and Chemical Engineering Department, Khalifa University, PO Box 127788, Abu Dhabi, United Arab Emirates

[4]Research and Innovation Center for graphene and 2D materials (RIC2D), Khalifa University, PO Box 127788, Abu Dhabi, United Arab Emirates

[5]Departement Fysica, Universiteit Antwerpen, Groenengogerlaan 171, 2020 Antwerp, Belgium

[6]Departamento de Física, Universidade Federal do Ceará, Fortaleza, Ceará 60455-760 Brazil

[7]School of Physics and Optoelectronic Engineering, Nanjing University of Information Science and Technology, Nanjing 210044, China.

**Abstract**

Graphene's $sp^2$-bonded carbon lattice makes it exceptionally resistant to covalent functionalization, with existing approaches relying on chemically specific and often harsh methods, fragmenting our understanding of its reactivity. Here we show that electric fields of approximately 1 V nm$^{-1}$ in non-aqueous electrolytes enable the electrochemical functionalization of monolayer graphene with a broad range of substituents, leading to two-order-of-magnitude conductance modulation stable over 5,000 cycles and semi-reversible electronic conductor-insulator transitions. Analysis of the oxidation potentials across substituents reveals trends governed by classical descriptors. Species forming localised $\sigma$ bonds display increasing oxidation potentials with the meta-Hammett constant ($\sigma_m$), whereas substituents interacting via delocalised $\pi$ orbitals follow an inverse dependence on the modified Swain-Lupton resonance parameter ($R$). These relationships indicate that graphene's oxidation potential increases with the electron-withdrawing strength of the substituent. Our results provide new insights into the electrochemical activity of inert materials under strong electric fields.

**Main**

Functionalising graphene requires a wide range of methods, depending on the chemical species, so-called substituents, involved. The most common graphene derivative, graphene oxide, is usually synthesised using concentrated sulfuric acid, strong oxidisers, and elevated temperatures[1-5]. While highly effective, this method is limited to oxygen-containing functional groups and is compatible mostly with chemically exfoliated graphene suspensions. Alternative chemistries typically require highly energetic species excited via plasmas[6-10], hot atom sources[11-16], electrochemical methods[17-19] or photo-driven processes[20-24]. These have enabled the functionalisation of graphene with substituents like halogens[9,10,24-26], organic cations[1,27,28], and hydrogen[13-16,29-31]. More recently, electrochemical studies have shown that strong electric fields in non-aqueous electrolytes can drive reversible hydrogen chemisorption on graphene[32-34], achieving a high density of adsorbed protons that induces a reversible electronic conductor-insulator transition stable for > $10^6$ hydrogenation-dehydrogenation cycles. The insulating phase is attributed to disorder introduced by the high density of carbon-hydrogen bonds, which act as electronic carrier scatterers[13,33,35]. This raises the broader question of whether strong electric fields can generally enable high-density chemisorption of substituents in graphene. This possibility remains largely unexplored because existing approaches cannot reach the required coverage to trigger the conductor-insulator transition, are irreversible, or are limited to specific substituents. Here, we show that graphene can be reversibly functionalised electrochemically with a wide range of substituents under strong electric fields, and that this modulates its electronic conductivity. For the highest density of substituents, this leads to semi-reversible electronic conductor-insulator transitions. Analysis of the oxidation potentials reveals that the oxidation potential is governed by the electron-withdrawing strength of the substituent.

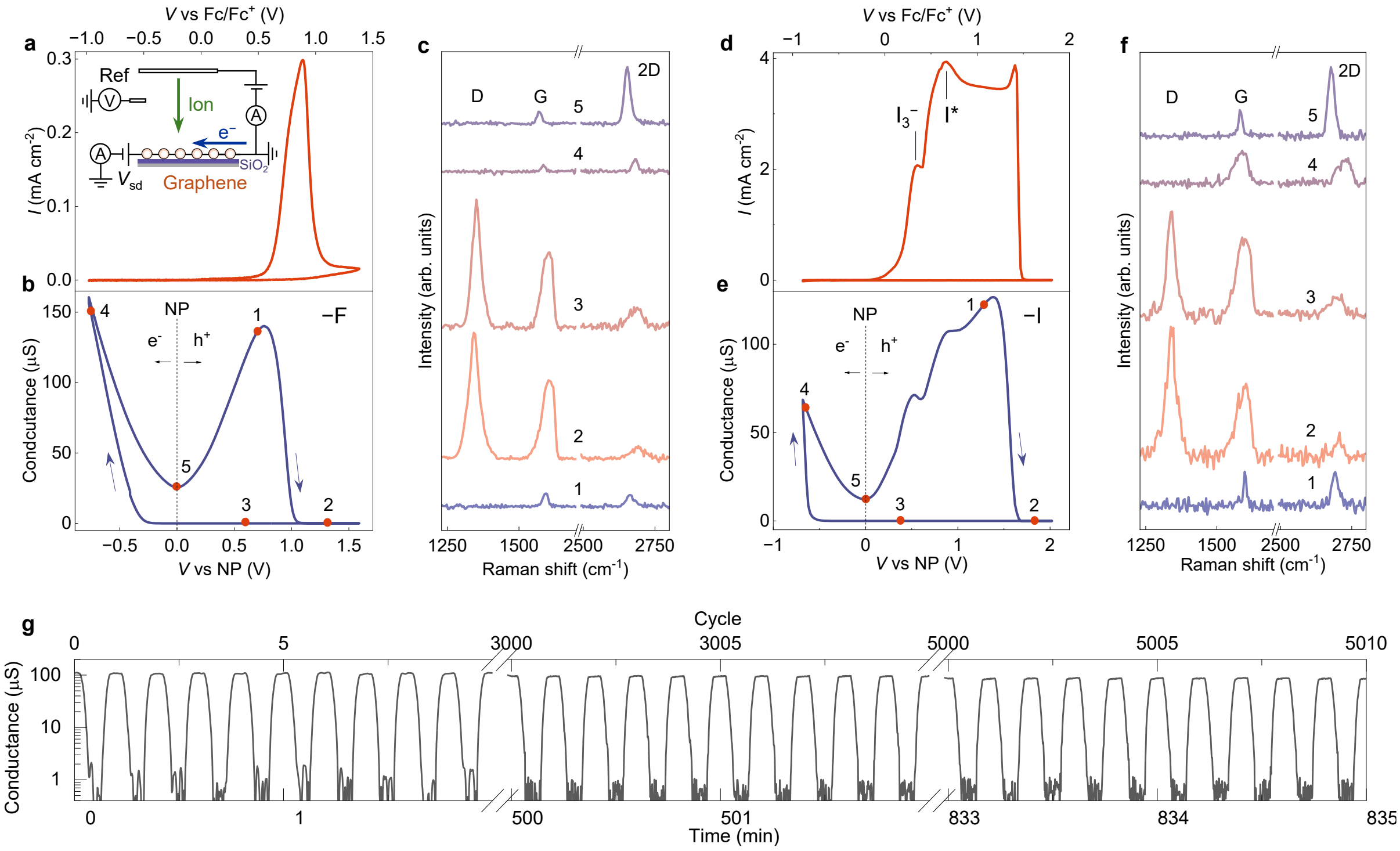

**Fig. 1 | Electrochemical functionalisation of graphene.** **a**, Electrochemical current density, *I*, as a function of potential vs Fc/Fc$^+$ for graphene electrode in fluoride electrolyte. Inset, experimental setup. **b**, In-plane electronic conductance vs potential relative to the neutrality point (NP) measured simultaneously with the electrochemical response. Graphene is electron- (hole-) doped at negative (positive) potentials vs NP. Numbered dots mark the potentials at which Raman spectra were acquired. **c**, Raman spectra obtained at the potentials marked with red dots in **b**. Each spectrum corresponds to a different potential. **d**, Electrochemical current density, *I*, as a function of potential vs Fc/Fc$^+$ for graphene exposed to iodide electrolyte. **e**, Electronic conductance for the iodide system measured simultaneously. Numbered dots indicate potentials at which Raman spectra were acquired. **f**, Raman spectra obtained at the potentials marked with red dots in **e**. **g**, Electronic conductance of graphene in fluoride electrolyte under a square-wave potential (0.8 V to -0.5 V vs NP, 10 s period), demonstrating 5,000 cycles of reversible conductance modulation over > 13 h.

**Experimental devices.** The devices for this work consisted of mechanically exfoliated graphene supported on $SiO_2$ substrates (Fig. 1a inset and Supplementary Fig. 1). The devices were coated with various non-aqueous electrolytes dissolved in polyethylene glycol, a solvent with a wide (> 4 V) electrochemical stability window[36], and electrically connected into the electrical circuit shown schematically in Fig. 1a. During the measurements, a voltage was applied to graphene vs the ferrocene/ferrocenium ($Fc/Fc^+$) redox couple (Supplementary Fig. 2 and Reference electrode in Methods). This voltage drives an electrochemical current in the device and simultaneously acts as a gate voltage that modulates graphene's in-plane electronic conductivity[32,33,37]. We also monitored the electronic current by applying a source-drain voltage. This allows referencing electrochemical processes to the Fermi energy of electrons in graphene[33] and enables the direct estimation of the electric field at the interface[33] (Estimation of electric field and Fermi energy in Methods).

Our main finding is that at electric fields of around 1 V $nm^{-1}$, corresponding to potentials > 0.9 V vs $Fc/Fc^+$, graphene's otherwise inert surface can be reversibly functionalised with a wide range of ionic species. All these reactions lead to conductor-insulator transitions accompanied by a sharp D band in the Raman spectra of the samples. Independent XPS analysis of the functionalised samples confirm the formation of covalent bonds with the substituents. We focused on compounds that could yield reversible chemisorption on graphene. Phenomenologically, the compounds can be grouped into two classes. The first produces relatively small electrochemical currents (peak current density of ~0.1 mA $cm^{-2}$) and exhibits simple, featureless response apart from a main redox peak in the electrochemical signal and the conductor-insulator transition in the electronic response. The second group generates currents approximately an order of magnitude larger (peak current density of ~1 mA $cm^{-2}$), and in some cases, the devices display additional features in both their electronic and electrochemical response beyond the main oxidation peak and the conductor-insulator transition.

**Compounds exhibiting a single redox process.** We begin discussing the species that functionalised graphene via a single-step electrochemical process. The species in this group were fluoride (−F), nitrate ($-ONO_2$), triflate ($-OSO_2CF_3$), boric acid (−O, BA), and Brønsted acids such as bis(trifluoromethane)-sulfonimide (HTFSI), triflic acid, and ammonium ($-HNH_3$). These species show no clear evidence of prior electrochemical transformation under the present conditions. Fig. 1a and b show the electrochemical and electronic response of a typical graphene device during oxidation measurements with potassium fluoride, as a representative example. The electronic response of graphene as a function of applied gate voltage is characterised by a minimum in conductance, known as the charge neutrality point (NP), which we found corresponds to -0.2 V vs $Fc/Fc^+$. For potentials positive vs the NP, graphene is hole-doped, and for a potential of 1.1 V vs NP (0.9 V vs $Fc/Fc^+$), we observed a sharp conductor-insulator transition in graphene's electronic response. This transition was accompanied by a sharp D band in the Raman spectrum of the device and the suppression of electrochemical activity (Fig. 1c, each spectrum was taken at a different potential). From the ratio of the D to G band intensities, the surface coverage of $sp^3$ bound fluorine can be estimated[38-40] at ~$10^{14}$ $cm^{-2}$, comparable to densities observed for electrochemically hydrogenated graphene[32-34]. This assignment is confirmed by independent XPS measurements using large area graphene samples grown via chemical vapour deposition, which confirm the presence of covalent C-F bonds ('XPS measurements' in Methods and Supplementary Fig. 3). Conversely, the electrochemical response (Fig. 1b) displayed an oxidation current with a peak current density of ~0.1 mA $cm^{-2}$, followed by the complete suppression of electrochemical activity upon reaching the insulating transition. Similar behaviour was observed for all species in this group, which displayed featureless oxidation responses with peak current densities of ~0.1 mA $cm^{-2}$ prior to the insulating transition ('Compounds exhibiting a single redox process' in Methods and Supplementary Fig. 4).

All the redox-induced conductor-insulator transitions reported in this work could be reversed across all compounds studied. However, the number of redox cycles that the samples endured depended on the applied potential protocol. Fig. 1g illustrates the case of fluorine. Setting the potential such that the conductance decreased by 100× yielded over 5,000 cycles of electronic conductance modulation with minimal conductivity loss over > 10 h of continuous measurement. However, increasing the potential, such that the full insulating transition was reached, led to device degradation within tens of cycles, as evidenced by a continuous degradation of electronic conductivity and the irreversible emergence of a D band in the Raman spectra (Supplementary Fig. 5). This was observed across all oxidation compounds, suggesting that reaching the insulating transition is the key driver of irreversible conductance modulation. Characterisation of such samples under optical and atomic force microscope (AFM) revealed no obvious cracks or macroscopic damage (Supplementary Fig. 6) and XPS measurements of similarly measured CVD samples confirmed the presence of C-F bonds (Supplementary Fig. 3), with a small contribution from C-$F_2$ bonds (indicative of functionalised vacancies) [41-43]. On this basis, we attribute the semi-reversible nature of the conductor–insulator transition in these samples to the incomplete removal of covalent bonds during subsequent cycles, together with the formation of a small density of lattice vacancies. The hydrogenation reaction was notably more robust, with devices enduring cycles indefinitely within our experimental timeframes, in agreement with previous reports[32-34], and consistent with hydrogen's light mass.

**Compounds exhibiting multiple redox processes.** The second group of species are more susceptible to electrochemical oxidation and are oxidised prior to the functionalisation of graphene. This results in notably larger electrochemical current densities, of the order of 1 mA $cm^{-2}$, and additional features in the electrochemical and electronic responses beyond the conductor-insulator transition. The species in this group are iodine (−I), nitrite (−ONO) and chloride (−Cl), see Supplementary Fig. 7. Fig. 1d and e show the electrochemical and electronic response of graphene during oxidation measurements with iodine ions as a representative example. Two local minima were apparent within the hole-doped branch of the electronic response at well-defined potentials (Fig. 1e) and were reproducible across multiple cycles (Supplementary Fig. 8). Such conductance minima occur when the Fermi level moves towards the neutrality point, reducing carrier density. Because the minima appear in the hole-doped branch, they indicate electron doping of graphene, consistent with oxidation of the substituent. Raman spectroscopy provides further insights. No D band was observed at these minima (Fig. 1f, each spectrum was taken at a different applied potential), confirming that the doping does not arise from covalent functionalisation. Hence, the substituents physisorb on the surface, leading to charge transfer without the formation of $sp^3$ bonds. Fig. 1d shows that the electrochemical current displayed local minima at the same potentials observed in the electronic response. Given previous work on the redox chemistry of iodine[21,44], we attribute the first minima (0.3 V vs $Fc/Fc^+$) to the formation of triiodide ($I_3^-$), and the second one (0.7 V vs $Fc/Fc^+$) to the appearance of the atomic iodine species ($I^*$)[21,44], as such peaks can also be observed in control experiments with Pt electrode in agreement with previous experiments[21,44] (Supplementary Fig. 8). Higher potentials (1.6 V vs NP, that is 1.4 V vs $Fc/Fc^+$) trigger a conductor-insulator transition in graphene accompanied by a strong D band in the Raman spectra and the suppression of electrochemical activity (Fig. 1f), confirming covalent functionalisation. This is confirmed with independent XPS characterisation (Supplementary Fig. 3). All species in this group produced similarly strong electrochemical responses, with peak currents on the order of 1 mA $cm^{-2}$, which we attribute to additional Faradaic steps besides graphene oxidation.

**Oxidation potential trends.** Fig. 2a and Supplementary Table 1 summarise the onset potentials for graphene functionalisation across all compounds studied. Each data point represents the average of typically 2-4 independent devices, with onset potentials varying by ≈ 0.1 V between different devices. Measurements performed using salts with different cations shifted the onset potentials by up to ≈ 0.2 V but did not alter the overall trends (Meta-Hammett and modified Swain–Lupton analysis, Supplementary Fig. 9). We present this trend both in terms of potential vs $Fc/Fc^+$ and the Fermi energy of electrons in graphene obtained from the electronic response, which serves as an independent reference. We also estimated the electric field at which these reactions take place, and found it increases with oxidation potential from 0.6 to 1.2 V $nm^{-1}$ (Estimation of electric field and Fermi energy in Methods). Despite these clear physical trends, the oxidation potential showed no consistent dependence on common descriptors such as substituent size, electronegativity, or oxidising strength. There was also no consistent trend based on whether the species exhibit single or multiple redox processes. Hence, while the compounds can be grouped electrochemically by their current response as discussed above, this classification does not elucidate any trends in oxidation potentials. This raises the question of whether a more general chemical principle governs graphene's functionalisation under strong electric fields.

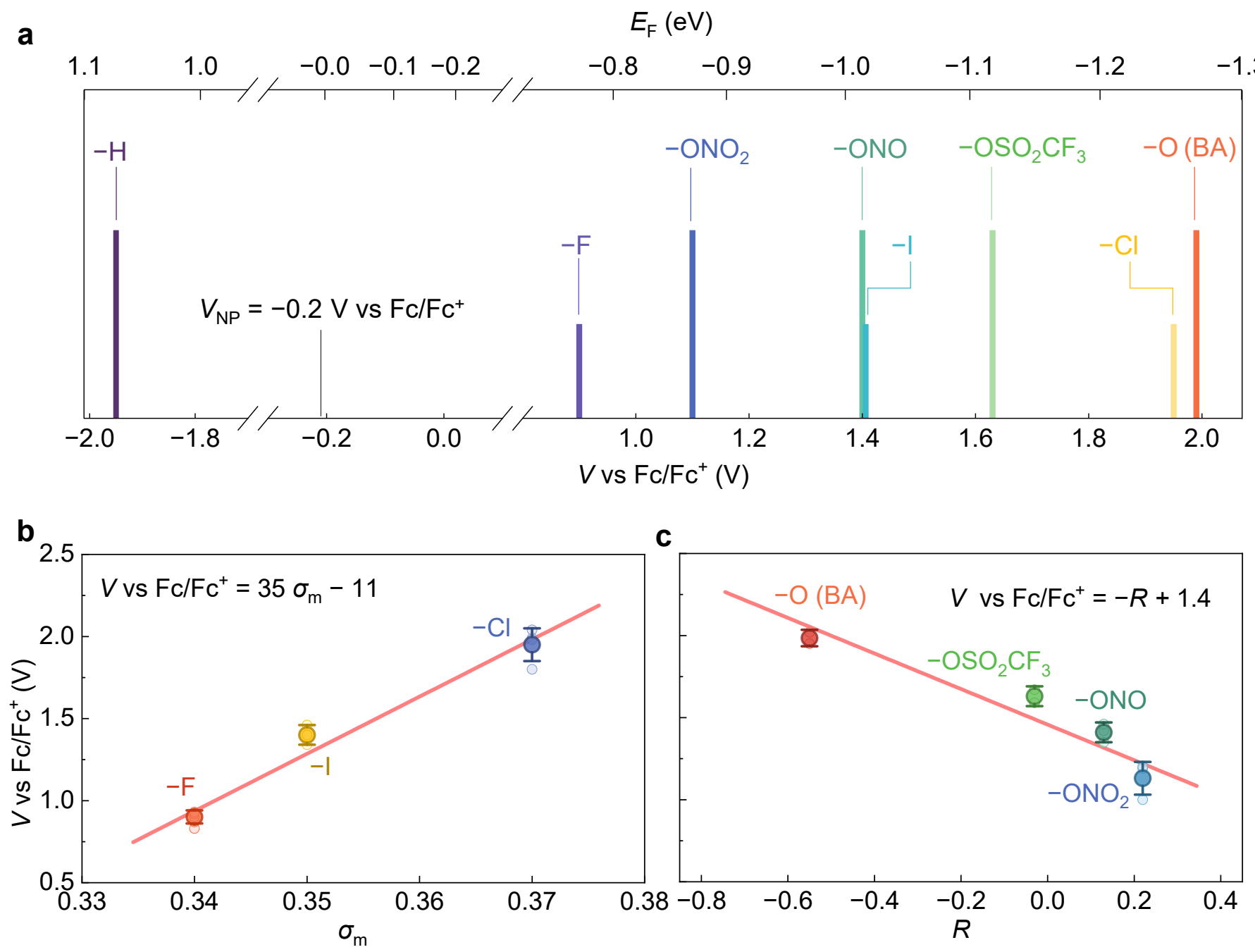

**Fig. 2 | Substituent-dependent trends in graphene's functionalisation. a**, Potential vs $Fc/Fc^+$ and the Fermi level ($E_F$) at which chemical functionalisation of graphene occurs for the compounds tested. The neutrality point (NP) corresponds to -0.2 V vs $Fc/Fc^+$. **b**, Oxidation potential vs $Fc/Fc^+$ of halogen substituents on graphene as a function of the classical meta-Hammett constant ($\sigma_m$). The best linear fit (red line) shows a positive correlation between the oxidation potential and $\sigma_m$ with $r^2 = 0.98$. **c**, Oxidation potential vs $Fc/Fc^+$ of oxygen-containing substituents on graphene as a function of modified Swain-Lupton resonance constant, *R*. The

linear fitting (red line) shows a negative correlation between potential and $R$ with $r^2 = 0.91$. In **b** and **c**, small data points represent oxidation potential from individual devices extracted from at least 10 repetitions, while large data points represent mean values. Error bars indicate standard deviations (SD) across independent devices (device numbers, $n = 5$ for –F, 3 for –I, 4 for –Cl, 2 for –O(BA), 2 for $–OSO_2CF_3$, 3 for –ONO and 3 for $–ONO_2$).

In search of a unifying principle, we analysed the oxidation potentials using the meta-Hammett and Swain-Lupton formalism[45,46]. This approach is well established for the prediction of electrochemical and catalytic properties of small organic molecules[46-50] and has also been applied to gold nanoparticles[51], but has not been explicitly demonstrated for reactions on the basal plane of graphene. In this framework, substituents modify the electronic structure of a parent compound, here the graphene lattice, through two mechanisms. The first is the so-called inductive/field effect, whereby electron density is withdrawn from or donated to the parent compound through $\sigma$ bonds or electrostatic polarisation (Fig. 2b). The meta-Hammett constant $\sigma_{\mathrm{m}}$ is dominated by this effect and is thus used to quantify it ($\sigma_{\mathrm{m}} > 0$ for electron-withdrawing and $\sigma_{\mathrm{m}} < 0$ for electron-donating substituents). The second is the so-called resonance effect, in which substituents with delocalised $\pi$-electrons donate or withdraw electron density through conjugation (Fig. 2c). This contribution is quantified by the modified Swain-Lupton resonance parameter $R$ ($R > 0$ for electron-donating and $R < 0$ for electron-withdrawing substituents).

Fig. 2b shows that the oxidation potential for halogen functionalisation of graphene scales linearly with the classical meta-Hammett constant, $\sigma_{\mathrm{m}}$[46], following the relationship $V$ vs $Fc/Fc^+ = 35\ \sigma_{\mathrm{m}} - 11$ ($r^2 = 0.98$). Conversely, Fig. 2c shows that the oxidation potentials with $\pi$-conjugated substituents (oxygen-containing compounds) decrease linearly with the modified Swain-Lupton resonance parameter $R$[46]. In this case, the relation is $V$ vs $Fc/Fc^+ = -R + 1.4$ ($r^2 = 0.91$). For comparison, we tested whether $\pi$-conjugated substituents correlate with $\sigma_{\mathrm{m}}$ or halogens with $R$ but found no meaningful correlation in either case (Supplementary Fig. 10). These results thus show that the oxidation potential of graphene increases with the electron-withdrawing strength of the substituent. For halogens this behaviour is governed by inductive/field effects, whereas for $\pi$-conjugated substituents it is governed by resonance effects. Our DFT calculations of representative oxidation reactions are consistent with these observations (Supplementary Fig. 11 and 12). Charge-density-difference, Bader charge, and highest occupied molecular orbital analyses show that strong electric fields drive depletion of $\pi$ electrons above the graphene plane and that this depletion is substantially more delocalised for substituents such as $–ONO_2$ and −ONO ($\pi$ interaction) than for substituents such as −F or −Cl ($\sigma$ interaction). Ab initio molecular dynamics (AIMD) simulations (Supplementary Fig. 13) confirm the bonding configurations and locations under the applied electric field, providing direct theoretical support for the proposed bonding motif.

**Discussion.** We demonstrate an electrochemical route to covalently functionalise monolayer graphene under strong electric fields with non-aqueous electrolytes, enabling chemisorption of halides, nitrites, acids and larger anions. All substituents induce a conductor-insulator transition, consistent with disorder-driven carrier scattering[13,33,35]. Mechanistically, the oxidation potential follows classical Hammett and Swain-Lupton relationships, separating substituents dominated by inductive/field effects from those governed by resonance. More broadly, we show that strong electric fields transform chemically inert graphene into chemisorption platform which can be understood using classical descriptors.

## Methods

**Materials.** Poly(ethylene glycol) (PEG, average molecular weight Mn of 600) was used as solvent. The electrolytes used were potassium fluoride (KF, 99.9%), methylammonium iodide ($CH_3NH_3I$, 99%), methylammonium chloride ($CH_3NH_3Cl$, 99%), Ammonium chloride ($NH_4Cl$, 99.5%), lithium nitrate ($LiNO_3$, 99.999%), Sodium nitrate ($NaNO_3$, 99%), sodium nitrite ($NaNO_2$, 99%), triflic acid ($HSO_3CF_3$, 99%), boric acid ($H_3BO_3$, 99.5%), bis(trifluoromethane)sulfonimide (HTFSI, 99%). For reference calibration, we used silver nitrate ($AgNO_3$, 99.999%), ferrocene ($Fc/Fc^+$, 98%), tetrabutylammonium tetrafluoroborate ($TBA^+BF_4^-$, 99%), acetonitrile (CAN, 99.8%), and non-aqueous Ag reference electrode. All materials were obtained from Sigma-Aldrich.

**Device fabrication and measurements.** Source and drain electrodes (Au/Cr) were deposited on $Si/SiO_2$ substrates using photolithography and electron beam evaporation. Mechanically exfoliated monolayer graphene crystals with a rectangular shape were transferred onto the substrates to form the conducting channel between these electrodes. An SU-8 epoxy washer with a 15-μm diameter hole was transferred over the source and drain electrodes leaving a mid-section of the graphene flake exposed (Supplementary Fig. 1). To prepare the electrolytes, various compounds (e.g. KF, $NaNO_2$, etc) were dried at 60 °C overnight prior to use, inside a glovebox. The PEG was degassed by three vacuum-argon cycles, followed by drying at 60 °C under vacuum and stirring overnight inside a glovebox. The dried compounds were then dissolved in PEG to prepare 0.2 M electrolyte solutions inside a glovebox. Approximately 15 μL of the prepared solutions were drop-cast onto the devices. Then the devices were placed inside a sealed chamber filled with argon gas for measurements.

**Electrical and electrochemical measurements.** To measure the electrochemical response, graphene was connected to a palladium counter electrode and an $Ag/Ag^+$ reference electrode, and the $Fc/Fc^+$ redox couple was used as an internal reference (see 'Reference electrode' below for further details). The response was controlled with a dual-channel Keithley 2636B sourcemeter in potentiostat mode. We swept the potential on the graphene working electrode vs the reference at a rate of 10 mV $s^{-1}$. A second Keithley source meter was used to apply the source-drain bias ($V_{sd}$) and measure the in-plane electronic current. Electrochemical potentials are reported as $V$ vs $Fc/Fc^+$, using the internal calibration described below.

**Reference electrode.** The experiments were carried out with a PEG solution containing 0.2 M solutes as the supporting electrolyte. The graphene electrode was used as the working electrode, palladium as the counter electrode, and an $Ag/Ag^+$ as the reference electrode with a double-junction salt bridge. The inner compartment of the salt bridge contained 0.01 M $AgNO_3$ and 0.1 M $TBA^+BF_4^-$ in CAN, while the outer compartment was filled with 0.2 M $TBA^+BF_4^-$ to connect the non-aqueous Ag reference electrode with the test solution. To account for potential drift, the $Ag/Ag^+$ reference electrode was calibrated against the ferrocene/ferrocenium ($Fc/Fc^+$) redox couple in the same electrolyte at the end of each run[52,53]. Cyclic voltammetry (CV) measurements were carried out with a Keithley sourcemeter configured in potentiostat mode. Current vs voltage measurements were performed at a scan rate of 10 mV $s^{-1}$ within a potential window of -0.2 V to 1.0 V vs $Ag/Ag^+$, enabling accurate evaluation of the redox processes. The response curve confirmed quasi-reversible behaviour with near-Nernstian peak separation, and the half-wave potential ($E_{1/2}$) of $Fc/Fc^+$ was experimentally determined to be 0.36 V vs $Ag/Ag^+$ under these conditions. The value was used to calibrate all electrode potentials according to $V$ vs $Fc/Fc^+$ = $V$ vs $Ag/Ag^+$ − 0.36 V.

**Raman spectroscopy.** For Raman measurements, devices were placed inside a Linkam chamber with an optical window. The spectra of devices were measured in situ as a function of applied potential using a WITec Alpha300R Raman spectrometer with a 514.5 nm laser. For each bias, the measurement was performed with a 50× objective, 600 l $mm^{-1}$ grating, accumulating time of 75 s over 4 cycles, and the laser power was 0.5 mW.

The data show that the insulating transition induced by the chemical reactions is accompanied by the emergence of a pronounced D band, a marked increase in the G peak intensity, and broadening of the 2D band-features characteristic of strong disorder in the graphene lattice[39]. The density of chemisorbed atoms was estimated using the intensity ratio of the D and G bands, $I_D/I_G$. We found that $I_D/I_G$ displayed little variability across substituents, typically ranging from 1.1 -1.55, whereas the ratio of their integrated areas was typically $A_D/A_G \approx 1.05$ - 1.9. This analysis yielded an estimated adatom density of 0.55 – 0.9 × $10^{14}$ $cm^{-2}$.

**XPS characterisation.** To obtain sufficient signal for XPS, large-area CVD graphene samples were prepared and driven into the fully functionalised state under electrochemical conditions equivalent to those used for the microdevices. Cu-grown CVD graphene (1 × 1 $cm^2$) was transferred onto Au-coated $SiO_2$ substrates using a PMMA-assisted method. Separate samples were intermittently biased at the corresponding functionalisation potential in KF, $CH_3NH_3I$, $CH_3NH_3Cl$, $NaNO_3$ or $NaNO_2$ for a cumulative 30 min, then thoroughly washed and dried before XPS analysis.

X-ray Photoelectron Spectroscopy (XPS) was performed using a Scienta Omicron ESCA2SR high-throughput system equipped with a monochromated Al Kα source ($h\nu$ = 1486 eV). High-resolution core-level spectra were acquired at pass energies of 50 eV. The spectra were corrected using the appropriate transmission functions, referenced to the graphitic $sp^2$ C–C peak at 284.8 eV and analysed using CasaXPS, shown in Supplementary Fig. 3.

The F 1*s* spectrum of fluorinated graphene showed a peak at 687.4 eV, assigned to covalently bound fluorine[42,54]. This value is higher than that typically reported for metal fluorides (684–685.5 eV) [55], and no K signal was detected after washing. The C 1*s* spectrum contained components at 285.8, 287.2, 288.3 and 289.8 eV, assigned to C–CF/$sp^3$, C–F, C–$CF_2$ and C–$F_2$ environments, respectively[41-43], supporting covalent fluorination rather than residual KF.

The I 3*d* spectrum contained I $3d_{5/2}$ components at 619.2 and 620.5 eV, assigned to ionic $I_3^-$ and covalent C–I species, respectively, with corresponding I $3d_{3/2}$ components at 630.8 and 632.1 eV[56,57]. The C 1*s* spectrum contained $sp^3$- and C–I-related components at approximately 285.3 and 286.0 eV, respectively[21,56,58].

Chlorinated graphene showed a Cl 2*p* spin–orbit doublet at 200.0 and 201.6 eV with an area ratio of approximately 2:1, consistent with covalently bound chlorine rather than ionic chloride[17,59]. The C 1*s* spectrum contained C–Cl- and C–$Cl_2$-related components at 287.3 and 289.9 eV, respectively.

The $NO_3^-$- and $NO_2^-$-treated samples showed O 1*s* peaks at approximately 533.7 eV, whereas only a negligible O 1*s* signal was observed for iodinated graphene measured under comparable conditions. Their C 1*s* spectra contained C–O, C=O and O–C=O components at approximately 287.3, 288.5 and 290.2 eV, respectively[60-62], consistent with the formation of covalently bonded oxygen-containing groups.

The proposed linkages are summarised in Supplementary Table 1 and Fig. 2b,c.

**Characterisation of samples after irreversible functionalisation**. Optical microscopy, AFM and Raman mapping were performed after repeated fluorination, following removal of the applied potential and washing of the electrolyte (Supplementary Fig. 5 and 6). The optical map shows no obvious tears or cracks, demonstrating macroscopic stability. AFM topography and adhesion maps, acquired using a Bruker Dimension Icon atomic force microscope, showed a pronounced increase in adhesion after repeated fluorination, whereas the corresponding height maps showed no obvious change (Supplementary Fig. 6). This indicates change in the surface chemistry. The Raman map shows the presence of a uniform D band throughout the sample, indicating the presence of $sp^3$ carbon. This could arise from either vacancies or to fluorine-carbon bonds that are not removed when the potential is reversed. XPS provides direct evidence for persistent covalent C–F and C–$F_2$ sites. These sites disrupt the conjugated $sp^2$ conducting network and are therefore the most plausible contributor to the incomplete recovery of the original conductance after the fully insulating state is reached. On this basis, covalently bonded fluorine atoms are the main driver of the samples' irreversible decrease in electronic conductivity.

**Compounds exhibiting a single redox process.** Besides fluorine, additional species from this group include nitrate (−$ONO_2$), which required a higher potential of ≈ 1.3 V vs NP (1.1 V vs $Fc/Fc^+$, Supplementary Fig. 4) to functionalise graphene. Triflate anion (−$OSO_2CF_3$) displayed a much higher potential of ≈ 1.8 V vs NP (1.6 V vs $Fc/Fc^+$, Supplementary Fig. 4); while the most demanding species was boric acid (−O, BA), which required ≈ 2.2 V vs NP (2 V vs $Fc/Fc^+$, Supplementary Fig. 4). This group also included protons, which stand out as the only species that functionalised graphene in the electron-doped branch of the response. Previous works[32,33] showed that HTFSI can hydrogenate graphene at ≈ -1.8 V vs NP (≈ -2 V vs $Fc/Fc^+$). We found that other acids, such as triflic acid ($HSO_3CF_3$) and ammonium (−$HNH_3$), hydrogenate graphene at the same potential (Supplementary Fig. 4). This similarity is consistent with a common proton-transfer event followed by graphene hydrogenation. Taken together, we attribute the moderate electrochemical current in this first group of species to their ability to directly functionalise graphene without intermediate redox processes.

**Compounds exhibiting multiple redox processes.** Besides iodine, additional species from this second group of compounds include nitrite (−ONO). During graphene's oxidation with this compound, we observed a distinct peak at ≈ 0.3 V vs $Fc/Fc^+$ (Supplementary Fig. 7), which can be attributed to the one-electron oxidation of $NO_2^-$ to form radical $NO_2$* [63,64], followed by the conductor-insulator transition at ≈ 1.6 V vs NP due to the adsorption of this radical (1.4 V vs $Fc/Fc^+$, Supplementary Fig. 7). Oxidation with chlorine ions followed a similar pattern (Supplementary Fig. 7). For this compound, we observed a weak shoulder at ≈ 1.3 V vs $Fc/Fc^+$, which can be attributed to the formation of Cl* radical[17] followed by the insulating transition at ≈ 2.2 V vs NP (2 V vs $Fc/Fc^+$, Supplementary Fig. 7).

**Estimation of electric field and Fermi energy.** For the electrolytes used here (0.2 M salt with a solvent dielectric constant of $\varepsilon_r \approx 10$) [36,65], the Debye length can be estimated as $\lambda_D \approx 0.3$ nm.

Because of this short screening length and the high gate potentials used, the electrical potential across the graphene-electrolyte interface drops almost entirely within the Stern layer. Consequently, each gate potential can be approximated using a parallel-plate capacitor model, which we employ to relate gate voltages, electric field $E$, and charge density $n$[66-70].

The charge induced by the gate is

$$neC^{-1} = (V - V^{NP}) \quad (1)$$

where the superscript NP marks the neutrality point and $e$ is the elementary charge constant. To consider the quantum capacitance of graphene, we note that

$$E_F = \hbar v_F \sqrt{\pi n} \quad (2)$$

with $v_F \approx 1 \times 10^6$ m s$^{-1}$ as the Fermi velocity in graphene and $\hbar$ the reduced Planck constant. This yields

$$V - V^{NP} = neC^{-1} + \hbar\, v_F\, e^{-1}(\pi n)^{1/2} \quad (3)$$

In previous work[33], we characterised the capacitance of similar electrolytes as $C \approx 20$ μF cm$^{-2}$. This yields $V - V^{NP} \approx 0.8 \times 10^{-14}$ V cm$^2$ $n + 1.16 \times 10^{-7}$ V cm $n^{1/2}$. Solving for $n$ provides the Fermi energy from the equation (2). To derive the relation between $V$ and $E$, we note that the electric field induced by the applied potential is

$$E = en\,(2\varepsilon)^{-1} = C\,(2\varepsilon)^{-1}\,(V - V^{NP}) \quad (4)$$

with $\varepsilon$ the dielectric constant of the solvent. This yields $E \approx 5.65 \times 10^8$ m$^{-1}$ $(V - V^{NP})$.

**Meta-Hammett and modified Swain–Lupton analysis.** The functionalisation potentials were analysed according to the classical meta-Hammett and modified Swain–Lupton formalisms[46]. Meta-Hammett substituent constants ($\sigma_m$) were used to analyse halogen substituents. Modified Swain-Lupton resonance parameters ($R$) were used to analyse $\pi$-conjugated substituents (oxygen-containing compounds). Linear regressions between the potentials and the corresponding substituent parameters were obtained using least-squares fitting. For reference, cross-analyses were also performed by correlating $\pi$-conjugated substituents with $\sigma_m$ parameters and halogen substituents with Swain-Lupton resonance parameters (Supplementary Fig. 10). $\pi$-conjugated substituents showed no correlation with $\sigma_m$, while halogens exhibited a non-linear trend, with increasing oxidation potential vs $R$. This latter result is chemically inconsistent, as higher $R$ implies stronger electron donation, which should lower rather than raise the oxidation potential. Control measurements were performed using salts containing the same anion but different cations, specifically $LiNO_3$ and $NaNO_3$ in the oxygen series, and $NH_4Cl$ and $CH_3NH_3Cl$ in the halogen series. The resulting onset potentials remained within ≈ 0.2 V of the corresponding average values, with the $LiNO_3$ and $NH_4Cl$ control measurements highlighted in red (Supplementary Fig. 9). These results thus confirm that halogen functionalisation is governed by inductive effects captured by $\sigma_m$, whereas resonance effects dominate the behaviour of $\pi$-conjugated substituents. The substituent parameters correspond to initial ions and were obtained from the reference[46].

**DFT calculations.** First principles calculations were carried out using the plane wave-based DFT method as implemented in the Vienna Ab Initio Simulation Package (VASP)[71]. Spin-polarized DFT calculations with a dispersion correction method (DFT-D3) were performed to account for van der Waals interactions of adsorption and functionalization in different systems. Long-range dispersion forces were obtained using Grimme's method[72]. The generalised gradient approximation with the Perdew-Burke-Ernzerhof functional (GGA-PBE) was used to calculate the energy[73]. Electron-ion interactions were described using the projector augmented wave (PAW) method[74]. An energy cut-off of 520 eV for the plane-wave basis set was used for the convergence of the total energy. The convergence criteria were set to $10^{-4}$ eV and 0.2 V $nm^{-1}$ for the electronic self-consistent iteration and the forces on each atom, respectively, while the Brillouin zone was sampled using Monkhorst-Pack mesh **k**-points with a reciprocal space resolution of $2\pi \times 0.04$ $Å^{-1}$. The graphene nanosheet with a single atomic layer was modelled using a 7 × 8 periodic supercell in the in-plane directions, and a vacuum spacing of 17.13 Å along the *z*-axis to form a cubic cell, since the potential correction is currently only implemented for cubic supercells in VASP for charged systems. The adsorbate was initially placed at the top adsorption sites of graphene, and the entire system was fully relaxed within the framework of spin-polarized DFT. For charged species, dipole corrections were included along the out-of-plane direction to correct the potential and force errors introduced by the periodic boundary conditions. To investigate the effect of external electric field on binding structures, calculations were carried out by applying a homogeneous external field ranging from -10 to +10 V $nm^{-1}$ along the out-of-plane (*z*) direction. For each field strength, the geometry of the adsorbate-graphene complex was fully relaxed and the corresponding bond lengths were extracted. In VASP, a negative sign of the applied field indicates that the electric field vector is oriented opposite to the cell axis (-*z*). The applied electric field in the calculations are intended to capture the qualitative trend of electric-field-induced modulation of graphene reactivity.
Following the DFT calculations, the electronic structure was further analysed by examining the spatial distribution of the highest occupied molecular orbital (HOMO) for the different functional groups (Supplementary Fig. 12). The HOMO isosurfaces reveal substituent-dependent differences in orbital coupling: halogen groups show increasing localization on the halogen atom from F to I, whereas oxygen-containing groups retain stronger contributions from the graphene $\pi$ network. These trends are consistent with the predominantly inductive effect of halogens and the enhanced resonance contribution of oxygen-containing substituents identified from the Hammett and modified Swain–Lupton analyses.

**AIMD simulations.** AIMD simulations were also carried out using VASP and performed in the canonical (NVT) ensemble using the Nosé-Hoover thermostat at 300 K with a time step of 1 fs. An external electric field was applied perpendicular to the slab (*z* direction) using the dipole correction method to eliminate artificial interactions between periodic images. Specifically, AIMD simulations for KF, $CH_3NH_3I$, and $NaNO_3$ were performed under an electric field of 10 V $nm^{-1}$ for 5 ps, whereas simulations for $NaNO_2$ and $NH_4Cl$ were carried out under 20 V $nm^{-1}$ for 5 ps. Electronic convergence was achieved with an energy criterion of $10^{-4}$ eV, and Gaussian smearing with a width of 0.05 eV was employed. These provide additional evidence for the proposed bonding motif and directly visualize the bond formation process under an external electric field.

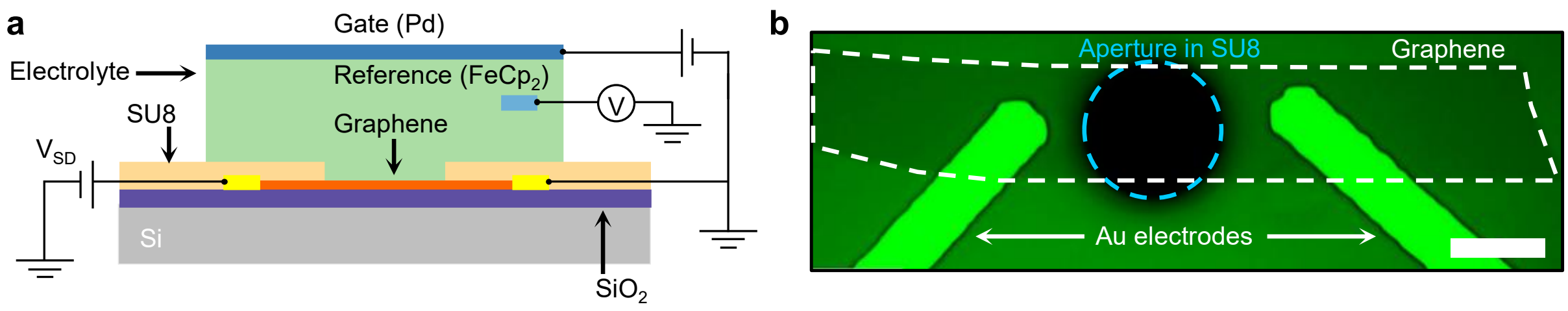

**Supplementary Fig. 1 | Experimental devices. a**, Schematic of experimental devices. **b**, Optical image of devices. Dashed white lines mark the graphene flake. Dashed blue circle (15 μm diameter), aperture in the SU-8 seal. Scale bar, 10 μm.

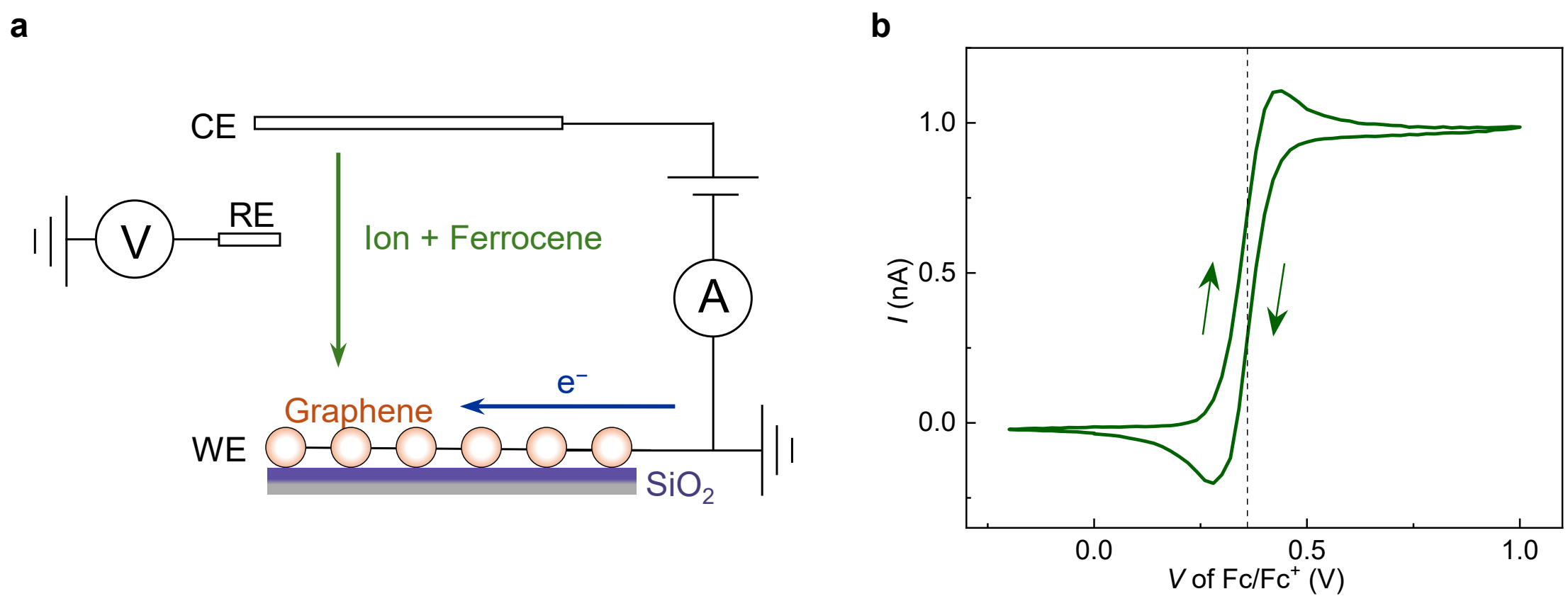

**Supplementary Fig. 2 | Calibration of Ag/Ag$^+$ reference electrode using ferrocene. a**, Experimental setup with graphene as the working electrode (WE), Ag/Ag$^+$ as the reference electrode (RE), and Pd as the counter electrode (CE). **b**, Cyclic voltammogram of ferrocene recorded in the electrolyte solution at a scan rate of 10 mV s$^{-1}$. Potentials are calibrated with the Fc/Fc$^+$ couple (half-wave potential $E_{1/2}$ = 0.36 V vs Ag$^+$/Ag under Ar marked by dashed line).

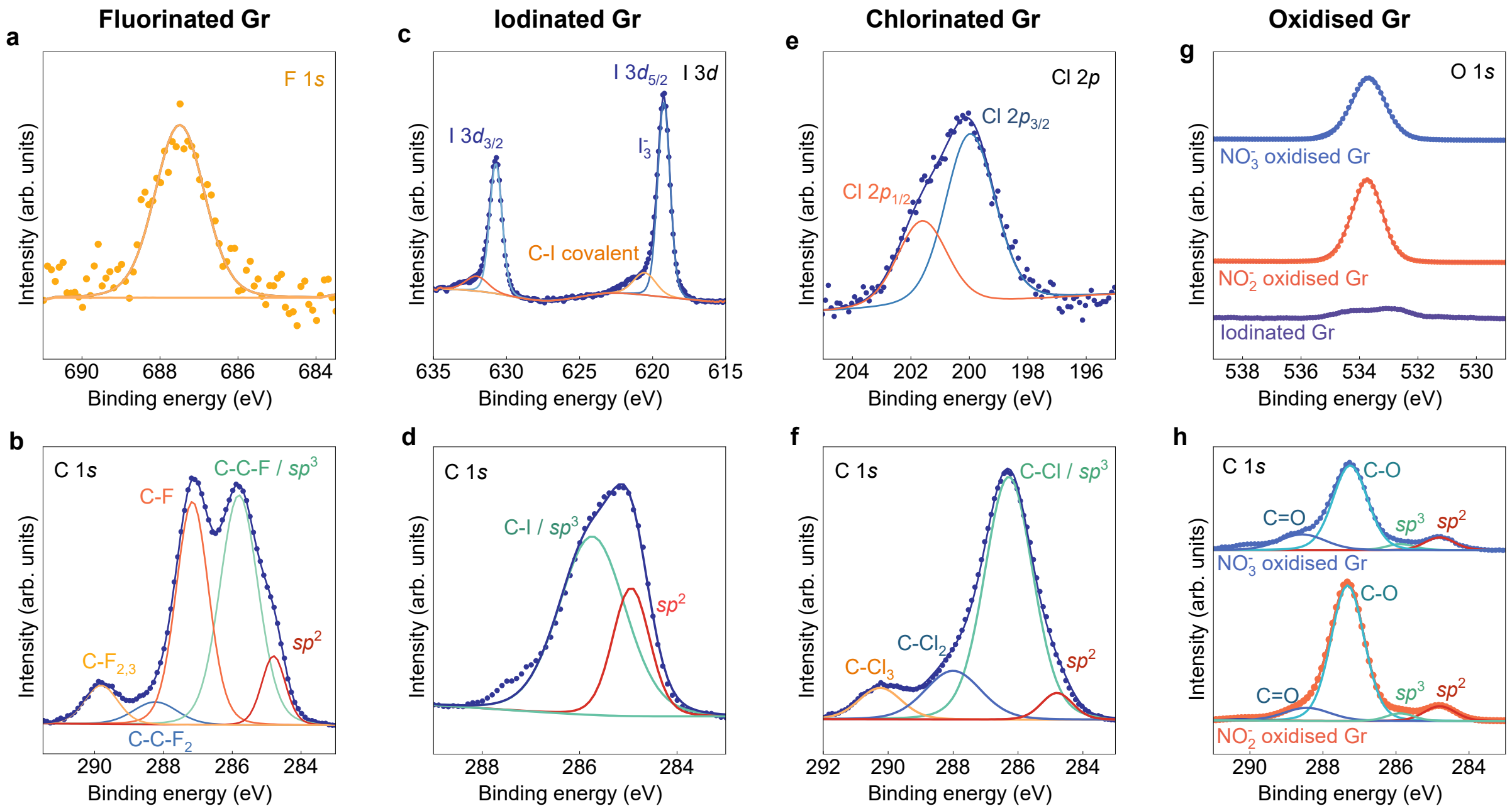

**Supplementary Fig. 3 | High-resolution XPS spectra of electrochemically functionalised CVD graphene after thorough washing and drying. a,b,** F 1*s* and C 1*s* spectra of fluorinated graphene. **c,d,** I 3*d* and C 1*s* spectra of iodinated graphene. **e,f,** Cl 2*p* and C 1*s* spectra of chlorinated graphene. **g,h,** O 1*s* and C 1*s* spectra of of $NO_3^-$- and $NO_2^-$-oxidised graphene, respectively. Scatters represent the experimental data, and solid curves show the fitted envelopes and individual components. The spectra in **g,h** are vertically offset for clarity.

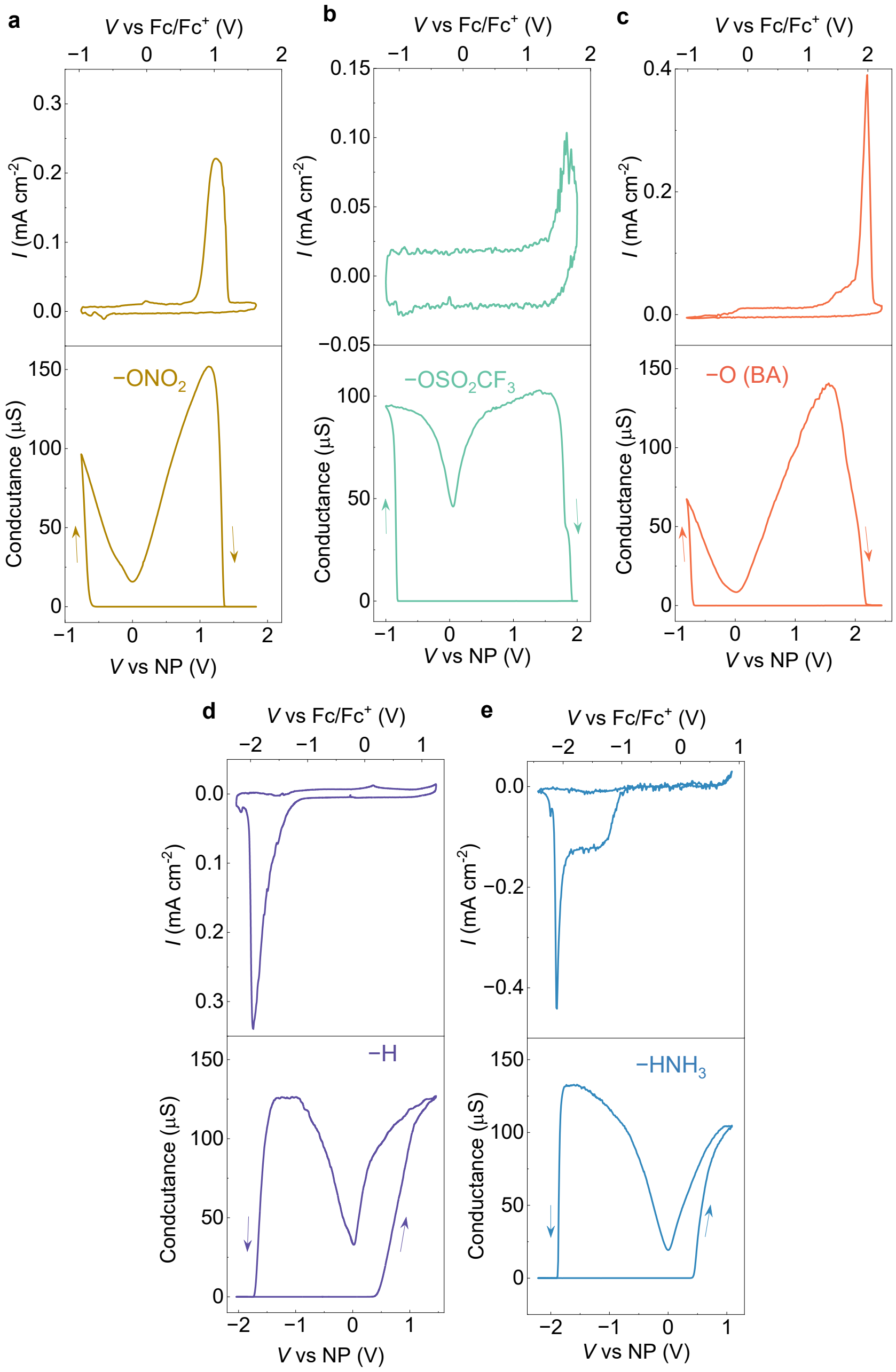

**Supplementary Fig. 4 | Direct electrochemical functionalisation of graphene with different substituents.** Electrochemical current density, *I*, (top) and electronic conductance (bottom) of graphene devices exposed to different oxidation electrolytes: **a**, nitrate, **b**, the triflate anion (–$OSO_2CF_3$) from triflic acid, **c**, boric acid, **d**, proton ($H^+$) from triflic acid and **e**, ammonium electrolyte, showing conductor–insulator transitions.

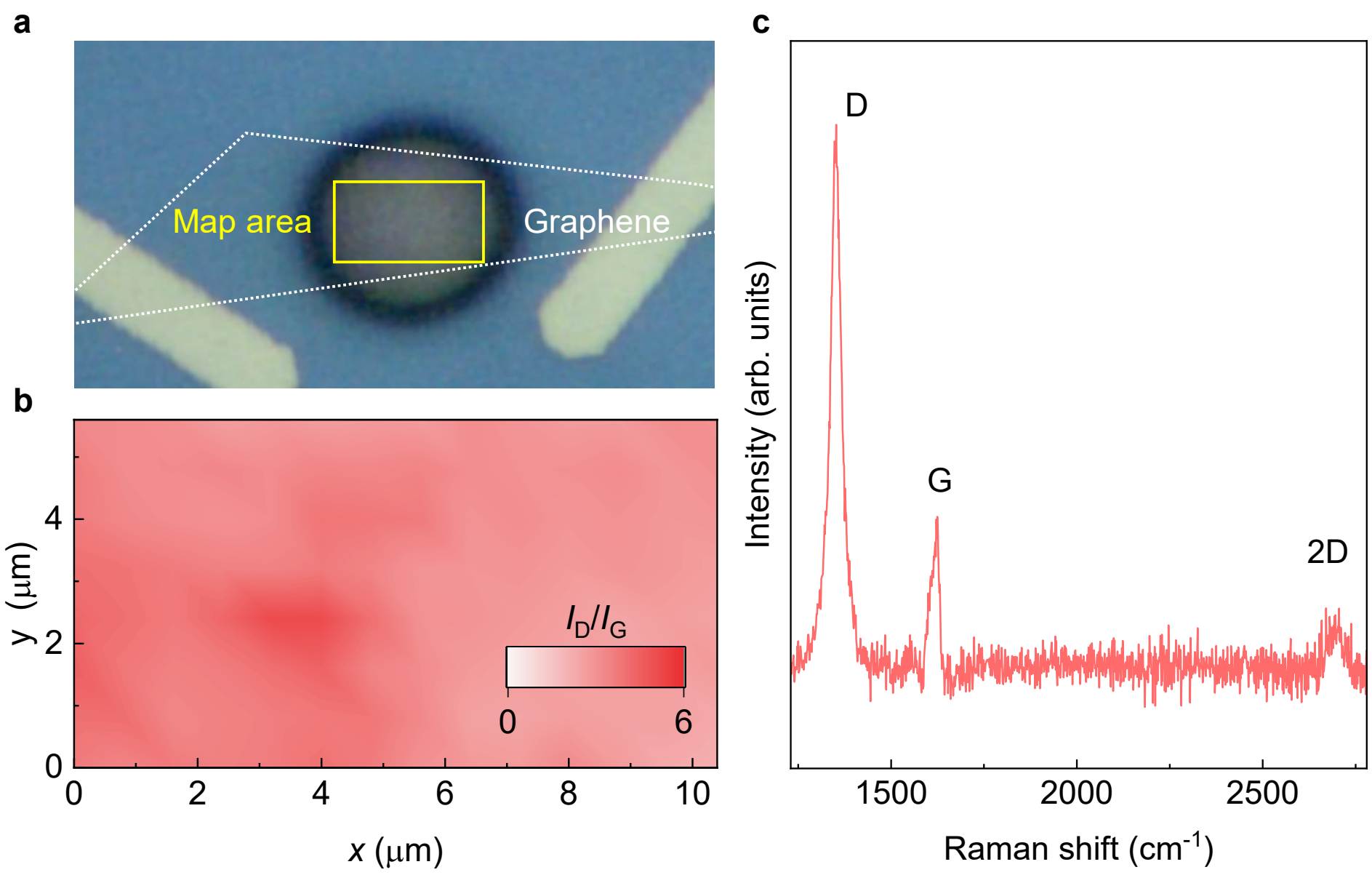

**Supplementary Fig. 5 | Raman characterisation of graphene after repeated fluorination. a,** Optical micrograph showing the graphene region outlined by the white dashed line and the Raman mapping area indicated by the yellow rectangle. **b, c,** Spatial maps of the integrated Raman intensities of the D band ($I_D$) and G band ($I_G$), respectively. **d,** Vertically offset Raman spectra acquired at all mapping positions.

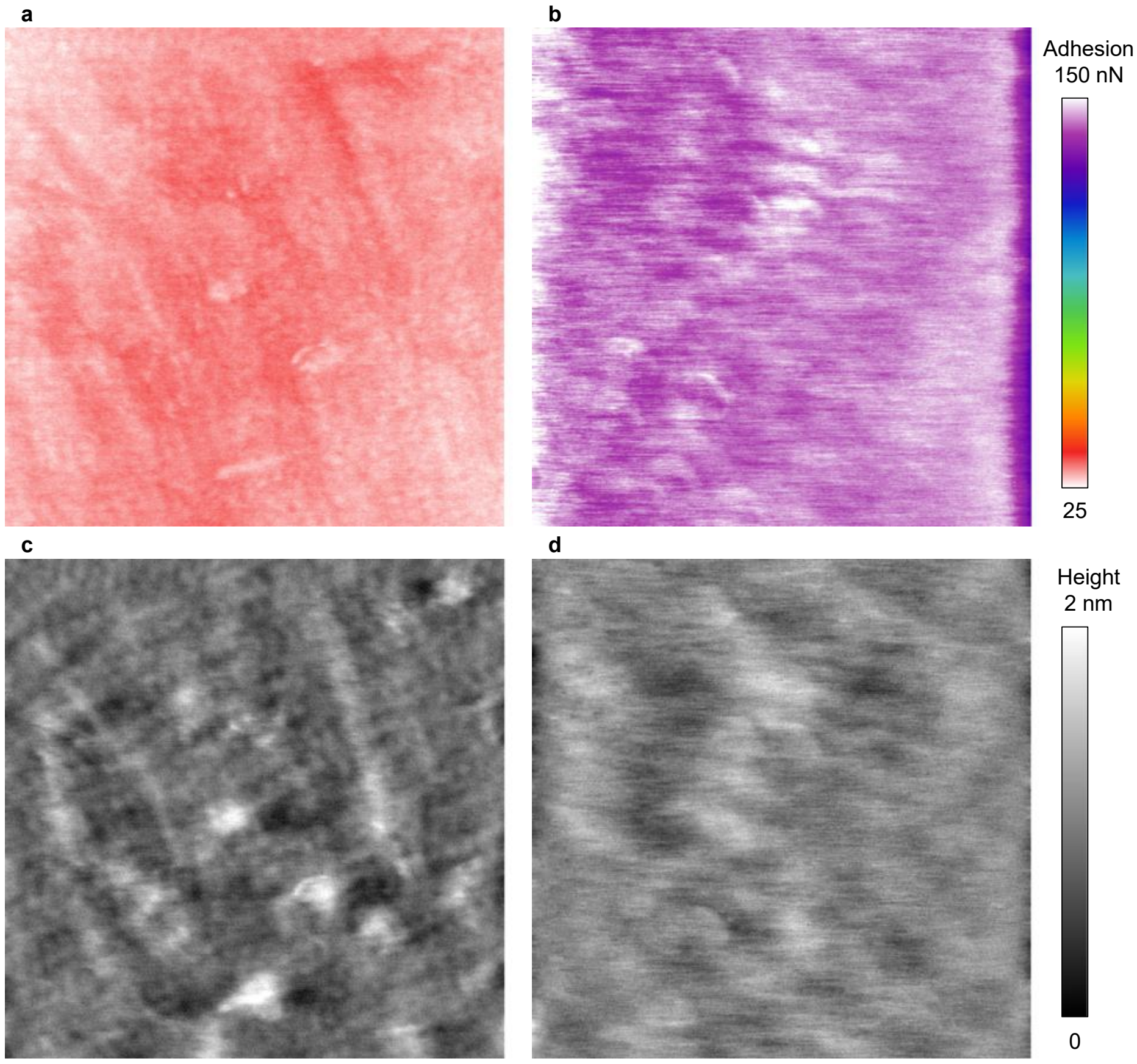

**Supplementary Fig. 6 | AFM characterisation of graphene before and after repeated fluorination. a, b,** PeakForce QNM adhesion maps of pristine graphene and graphene after repeated fluorination, respectively. **c, d,** Corresponding height maps. All images were acquired over 1×1 μm$^2$ areas using the same probe and measurement conditions. The adhesion and height maps are displayed using common scales of 25–150 nN and 0–2 nm, respectively.

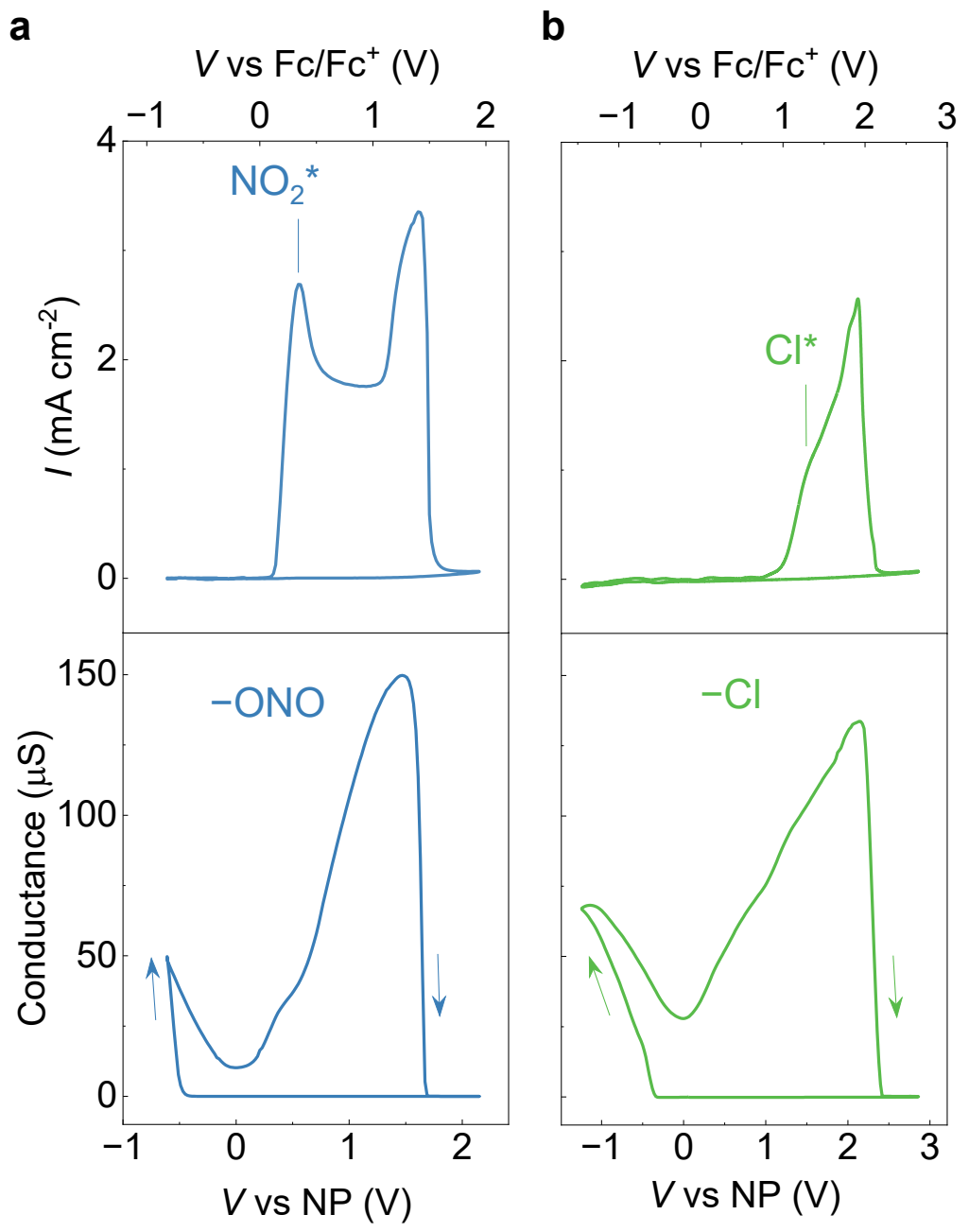

**Supplementary Fig. 7 | Electrochemical oxidation of graphene with compounds exhibiting multiple redox processes**. Electrochemical current density, *I*, (top) and electronic conductance (bottom) of graphene devices exposed to **a**, nitrite, and **b**, chloride electrolytes, showing conductor-insulator transitions.

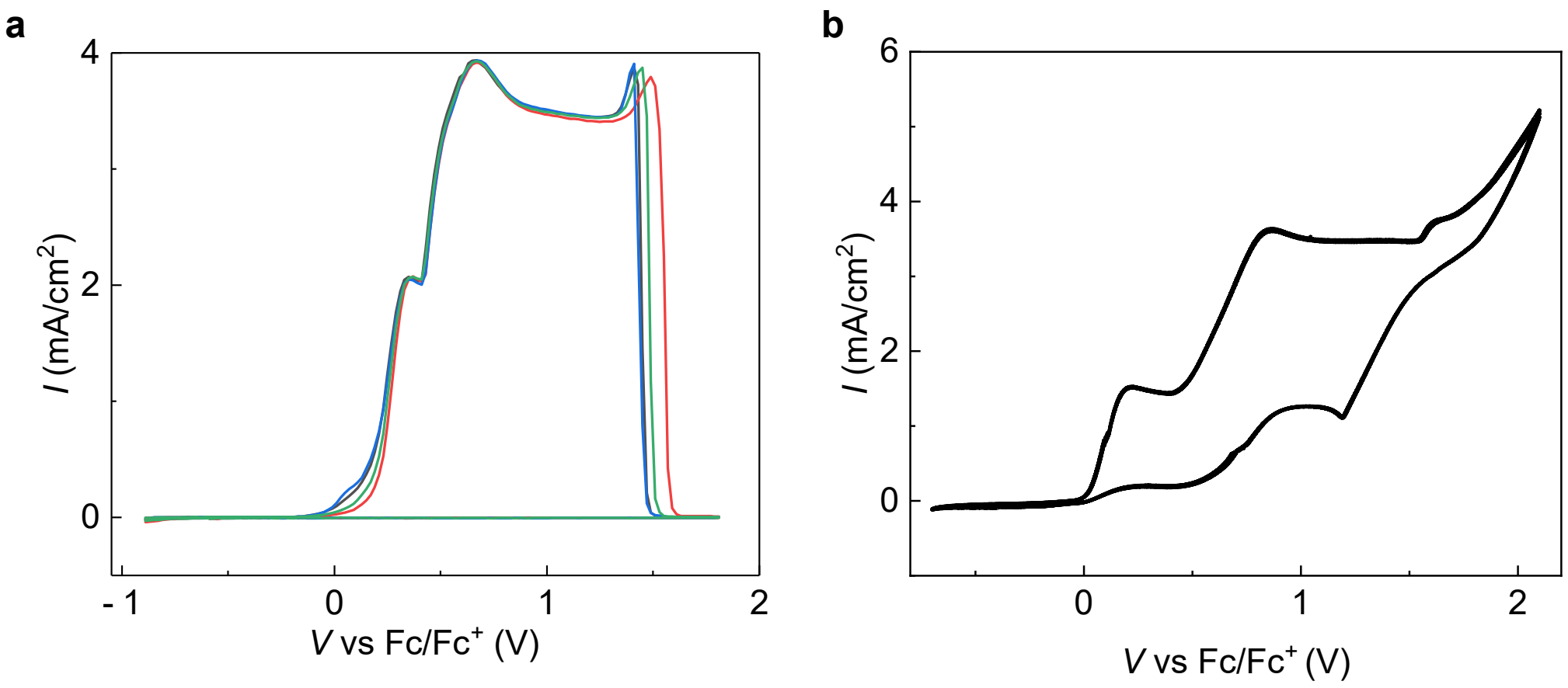

**Supplementary Fig. 8 | Electrochemical response of a, graphene and b, Pt in iodide electrolyte**. **a,** Electrochemical current density for graphene as a function of potential vs $Fc/Fc^+$ retraces itself in each cycle (marked with different colours). **b,** The Pt electrode displays increasing Faradaic current at potentials where the graphene current is suppressed.

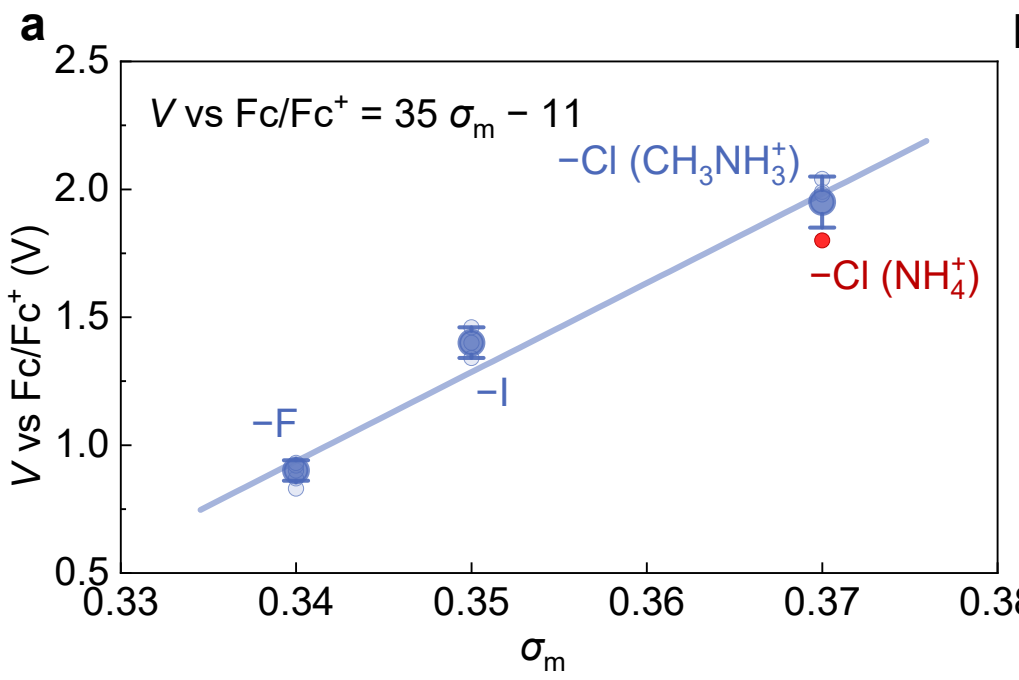

**Supplementary Fig. 9 | Effect of counter-cation identity on graphene functionalisation potentials. a**, Functionalisation potentials are shown for chloride salts containing $CH_3NH_3^+$ and $NH_4^+$ in the halogen series and **b**, nitrate salts containing $Na^+$ and $Li^+$ in the oxygen series. The $NH_4^+$ and $Li^+$ measurements are highlighted in red. In both cases, the onset potentials remain within ≈ 0.2 V of the corresponding average values and do not alter the observed correlations with the meta-Hammett $\sigma_m$ or modified Swain–Lupton resonance parameter $R$. Small data points represent oxidation potential from individual devices extracted from at least 10 repetitions, while large data points represent mean values. Error bars indicate standard deviations (SD) across independent devices (device numbers, $n$ = 5 for –F, 3 for –I, 4 for –Cl, 2 for –O(BA), 2 for $–OSO_2CF_3$, 3 for –ONO and 3 for $–ONO_2$).

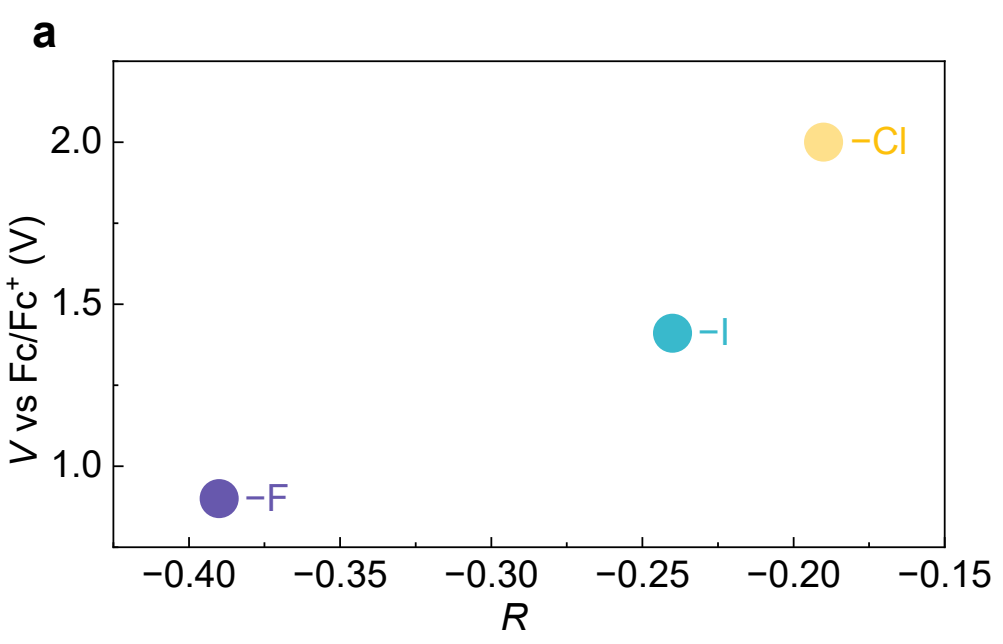

**Supplementary Fig. 10 | No meaningful correlation between halogen substituents and resonance effects, nor between π-conjugated substituents and inductive effects.** The measured oxidation potential vs $Fc/Fc^+$ of **a**, halogens on graphene as a function of the modified Swain-Lupton resonance parameter ($R$), and that of **b**, O- and N- substituents on graphene as a function of the meta-Hammett constant ($\sigma_m$). No correlation is observed between $\pi$-conjugated substituents oxidation potential and $\sigma_m$, whereas halogens display a chemically inconsistent increase of oxidation potential with $R$.

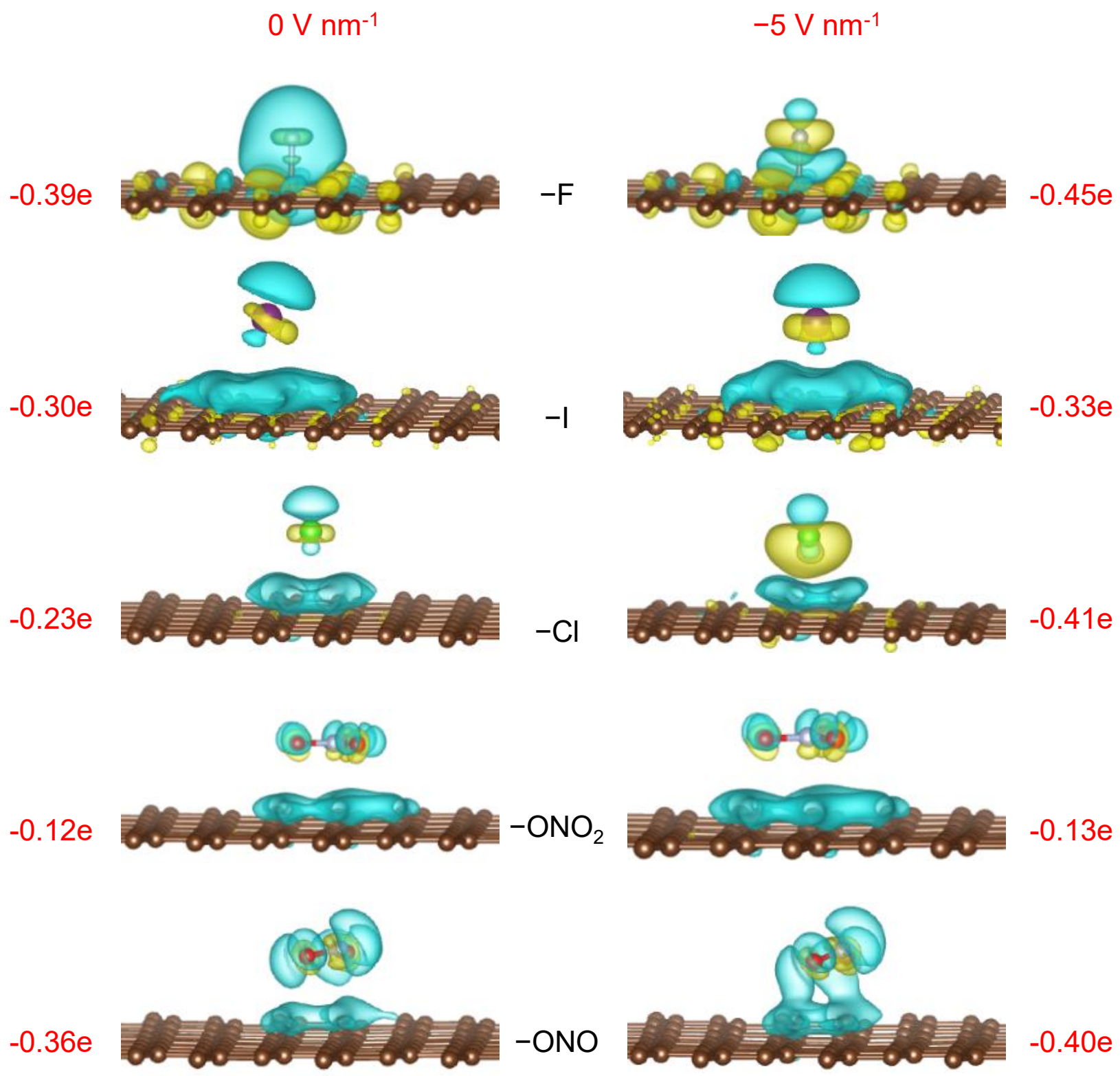

**Supplementary Fig. 11 | Calculated charge distribution during graphene functionalisation.** Left; 0 V $nm^{-1}$; right: -5 V $nm^{-1}$ applied electric field. Charge-density-difference distribution for different adsorbates on top of the graphene lattice (brown). The blue regions indicate electron depletion and the yellow regions electron accumulation. Larger isosurfaces of the charge density difference at the adsorbate graphene interface indicate stronger orbital hybridisation and bonding. The isosurface area generally increases with the applied electric field. The numbers on the left- and right-hand side of the figures represent the Bader charge transfer quantities for each case.

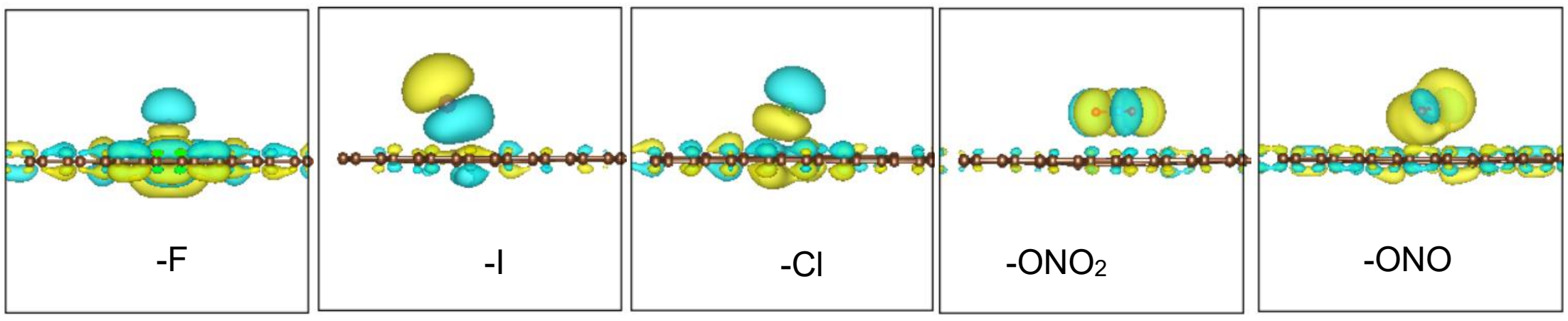

**Supplementary Fig. 12 | HOMO isosurfaces of graphene functionalized with different substituents under a homogeneous external field of -5 V nm$^{-1}$.** Spatial distribution of the HOMO depends strongly on the nature of the substituent. The yellow and green isosurfaces represent the positive and negative phases of the HOMO wavefunction, respectively. For the halogen-functionalized systems (–F, –Cl and –I), the HOMO contains contributions from both the graphene $\pi$ network and the newly formed C–X $\sigma$ bond. From F to I, the frontier orbital becomes progressively more localized on the halogen atom, reflecting the decreasing overlap between the graphene carbon and halogen *p* orbitals. In contrast, the oxygen-containing substituents (–$ONO_2$ and –ONO) retain a larger contribution from the graphene $\pi$ framework, indicating stronger $\pi$-mediated electronic coupling with graphene. These orbital distributions are consistent with the Hammett and modified Swain–Lupton analyses, in which halogens predominantly influence graphene through inductive effects, whereas oxygen-containing substituents exhibit a larger resonance contribution.

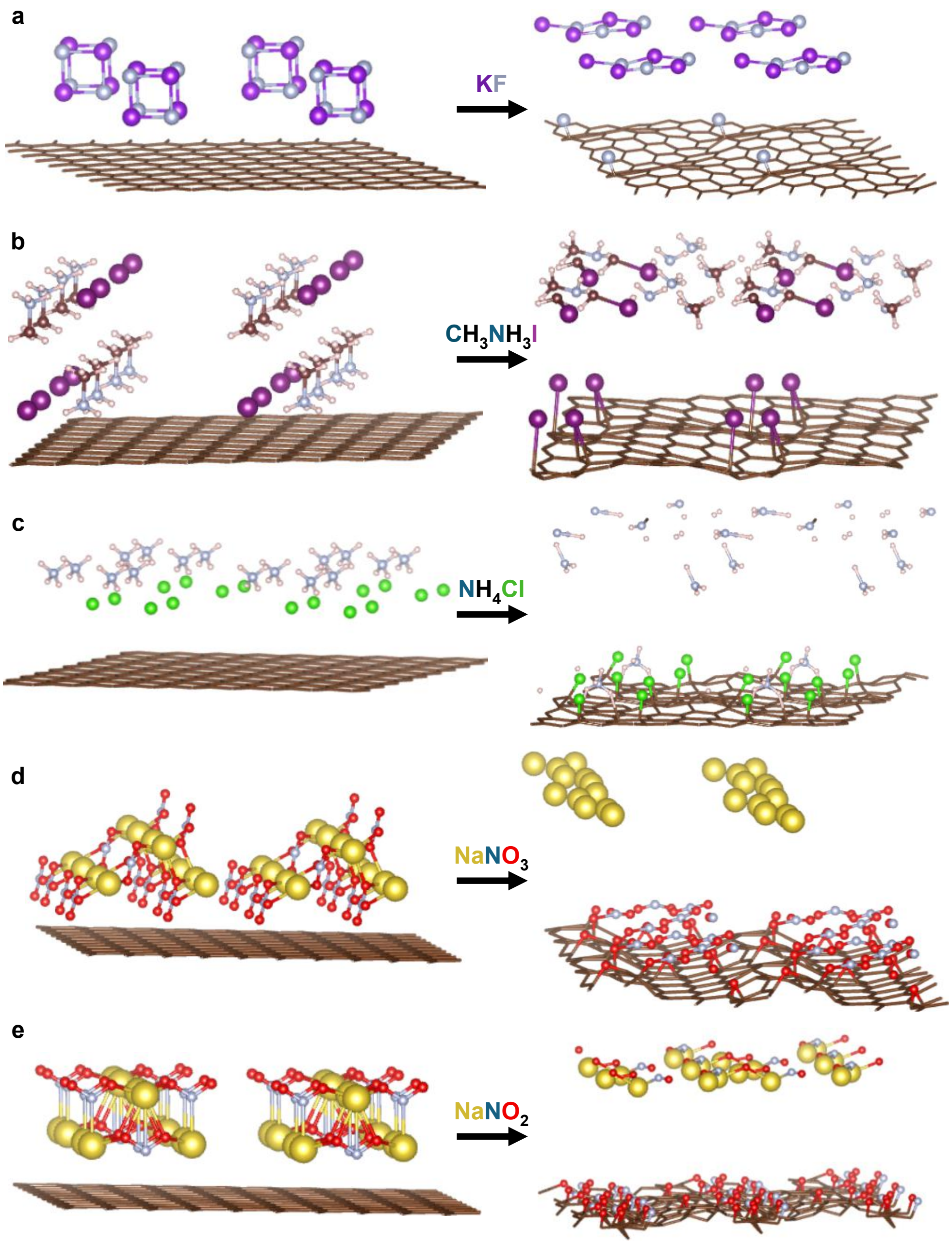

**Supplementary Fig. 13 | AIMD snapshots of electric field-induced bond between graphene and representative salts a, KF, b, $CH_3NH_3I$, c, $NH_4Cl$, d, $NaNO_3$, e, $NaNO_2$.** The left panels show the initial simulation configurations, while the right panels show the corresponding configurations after 5 ps of AIMD simulations under the applied electric field. Atoms are coloured according to their elemental identities.

**Supplementary Table 1: Potential for chemical functionalisation of graphene in various representative compounds.**

| Substituent | Proposed bonding motif | Chemisorption potential $V$ vs Fc/Fc$^+$ (V) | $V$ vs NP (V) | Meta-Hammett and Modified Swain-Lupton resonance Constants[46] |
|---|---|---|---|---|
| | | Oxidation | | |
| | Field/inductive contributions dominate | | | $\sigma_m$ |
| −F | C-F | 0.9 ± 0.04 | 1.11 ± 0.04 | 0.34 |
| −I | C-I | 1.4 ± 0.06 | 1.61 ± 0.06 | 0.35 |
| −Cl | C-Cl | 1.95 ± 0.1 | 2.16 ± 0.1 | 0.37 |
| | Resonance contributions dominate | | | $R$ |
| $-ONO_2$ | C-O | 1.1 ± 0.1 | 1.31 ± 0.1 | 0.22 |
| −ONO | C-O | 1.4 ± 0.06 | 1.61 ± 0.06 | 0.13 |
| $-OSO_2CF_3$ | C-O | 1.63 ± 0.06 | 1.84 ± 0.06 | -0.03 |
| −O (BA) | C-O | 1.99 ± 0.05 | 2.2 ± 0.05 | -0.55 |
| | | Hydrogenation | | |
| −H | C-H | -1.96 ± 0.08 | -1.75 ± 0.08 | - |